\documentclass[preprint]{aastex}

\begin{document}
\title{Detection of Lead in the Carbon-Rich, Very Metal-Poor Star LP~625-44: \\
A Strong Constraint on {\it s}-Process Nucleosynthesis at Low Metallicity}

\author{Wako Aoki\altaffilmark{1}, John E. Norris\altaffilmark{2}, Sean G. Ryan\altaffilmark{3}, Timothy C. Beers\altaffilmark{4}, Hiroyasu Ando\altaffilmark{1}}
\altaffiltext{1}{National Astronomical Observatory, Mitaka, Tokyo, 181-8588 Japan; email: aoki.wako@nao.ac.jp, ando@optik.mtk.nao.ac.jp}
\altaffiltext{2}{Research School of Astronomy and Astrophysics, The Australian National University, Private Bag, Weston Creek Post Office, Canberra, ACT 2611, Australia; email: jen@mso.anu.edu.au}
\altaffiltext{3}{Department of Physics and Astronomy, The Open University, Walton Hall, Milton Keynes, MK7 6AA, UK; email: s.g.ryan@open.ac.uk}
\altaffiltext{4}{Department of Physics and Astronomy, Michigan State University, East Lansing, MI 48824-1116; email: beers@pa.msu.edu}

\begin{abstract} 

We report the detection of the Pb {\small I} 4057.8\AA \ line in the very
metal-poor ([Fe/H]$=-2.7$), carbon-rich star, LP~625-44.  We determine the
abundance of Pb ([Pb/Fe] $ = 2.65$) and 15 other neutron-capture elements.  The
abundance pattern between Ba and Pb agrees well with a scaled solar system {\it
s}-process component, while the lighter elements (Sr-Zr) are less abundant than
Ba. The enhancement of {\it s}-process elements is interpreted as a result of
mass transfer in a binary system from a previous AGB companion, an
interpretation strongly supported by radial velocity variations of this system.

The detection of Pb makes it possible, for the first time, to compare model
predictions of {\it s}-process nucleosynthesis in AGB stars with observations
of elements between Sr and Pb.  The Pb abundance is significantly {\it lower}
than the prediction of recent models \citep[e.g., ][]{gallino98}, which
succeeded in explaining the metallicity dependence of the abundance ratios of
light {\it s}-elements (Sr-Zr) to heavy ones (Ba-Dy) found in previously
observed {\it s}-process-enhanced stars.  This suggests that one should either
(a) reconsider the underlying assumptions concerning the $^{13}$C-rich {\it
s}-processing site ($^{13}$C-pocket) in the present models,  or (b) investigate
alternative sites of {\it s}-process nucleosynthesis in very metal-poor AGB
stars.

\end{abstract}

\keywords{nuclear reactions, nucleosynthesis -- stars: abundances -- stars: AGB and post-AGB -- stars: carbon -- stars: Population II}

\section{Introduction}\label{sec:intro}

The slow neutron-capture process ({\it s}-process) is considered one of the
major pathways for the creation of nuclei heavier than iron, and the asymptotic
giant-branch (AGB) phase of low- and intermediate-mass stars has been studied
as its most likely astrophysical site.  One important component in
understanding {\it s}-process nucleosynthesis is the correct
identification of the neutron sources involved.  Two reactions --
$^{22}$Ne($\alpha,n$)$^{25}$Mg and $^{13}$C($\alpha,n$)$^{16}$O -- have
received most attention.  Recent models of AGB stars prefer $^{13}$C as the
main source, because the temperature of the He burning shell hardly reaches
$3\times 10^{8}$ K required for the $^{22}$Ne($\alpha,n$)$^{25}$Mg reaction
\citep[e.g., ][]{gallino98}. This is supported by the observed metallicity
dependence of the abundance ratios of heavier {\it s}-process elements (e.g.,
Ba, Nd) to lighter ones (e.g., Sr, Zr) found in {\it s}-process-enhanced
objects such as MS- and S-type stars \citep{smith90}, barium stars
\citep{luck91} and CH stars \citep{vanture92, norris97a}. While the seed nuclei
for the {\it s}-process, such as iron, are secondary (i.e., their abundances
are proportional to metallicity), the production of $^{13}$C in AGB stars is
primary, contrary to that of $^{22}$Ne.  Therefore, higher neutron exposure,
and thus larger enhancement of the heavier elements, is expected from $^{13}$C
in stars of lower metallicity.  Models of nucleosynthesis in AGB stars by
\citet{gallino98}, followed by \citet{busso99}, successfully reproduced the
observed trend for lighter elements (Sr-Zr), as well as for heavier ones
(Ba-Gd).

For stars of very low metallicity, according to these models, a
large excess of lead (Pb) is expected.  For instance, the enhancement of Pb by
two or three orders of magnitude relative to that expected for solar-abundance
stars is predicted for AGB stars with [Fe/H] $=-2.0$
\footnote{[A/B]$\equiv\log(N_{\rm A}/N_{\rm B})$ $-\log(N_{\rm A}/N_{\rm
B})_{\odot}$, and $\log\epsilon_{\rm A}$ $\equiv\log(N_{\rm A}/N_{\rm H})+12$
for elements A and B}, while that of Ba is at most one order of magnitude
\citep{busso99}. Thus Pb in metal-poor, {\it s}-process-enhanced, stars should
provide an excellent diagnostic for models of {\it s}-process nucleosynthesis
in AGB stars.

However, the abundance of Pb is difficult to measure in most stars.

Lead abundances for the metal-poor stars HD~115444 and HD~126238 were derived
from {\it Hubble Space Telescope} ultraviolet spectra \citep{sneden98}, but
Pb has not yet been detected in the optical spectra of these
objects.  The \citet{sneden00} analysis of a high-S/N Keck HIRES spectrum of
the {\it r}-process-rich star CS~22892-052 detected Pb {\small I} lines in the
visual spectrum of this star, and derived its abundance. We note
that the Pb observed in all three of these stars is attributed primarily to the
{\it r}-, rather than the {\it s}-process, due to the strong enhancements of
other {\it r}-process-dominated nuclei, such as Eu.  One study of the {\it
s}-process for a solar metallicity star by \citet{gonzalez98} reports the Pb
abundance of the post-AGB star FG-Sge, based on an analysis of the Pb {\small
I} 7229\AA \ line.

In this {\it Letter} we report the detection of Pb {\small I} 4057.8\AA \ and
derive a Pb abundance in the carbon-rich, very metal-poor star LP~625-44.  This
object was shown by \citet{norris97a} to exhibit very large excesses of carbon,
nitrogen, and neutron-capture elements.  Their interpretation was that the
large excesses of heavy elements were likely to have originated from {\it
s}-process nucleosynthesis in an AGB binary companion which provided LP~625-44
with carbon-rich material by mass transfer.  The updated abundance pattern (see
\S \ref{sec:ana}), and variation of radial velocity (see \S \ref{sec:obs}),
reported in the present work make this interpretation quite convincing.  The
determination of a Pb abundance for this star (\S \ref{sec:ana}) provides
the opportunity, for the first time, to test models of nucleosynthesis in AGB
stars for {\it s}-process elements between Sr and Pb (\S \ref{sec:disc}).

\section{Observations and Measurements}\label{sec:obs}

A high-resolution spectrum of LP~625-44 was obtained with the University
College London coud\'e \'echelle spectrograph (UCLES) and Tektronix
1024$\times$1024 CCD at the Anglo-Australian Telescope on August 5, 1996.  The
wavelength range 3700--4720 \AA \ was covered with resolving power $R \sim
40,000$.  We also obtained a red spectrum (5015--8500 \AA) with the same
instrument on June 16, 1994.  The numbers of detected photons are 12000 per
0.04\AA \ pixel at 4300\AA \ ($S/N \sim 150$ per resolution element) and 2000
per 0.06\AA \ pixel at 6000\AA \ ($S/N \sim 60$ per resolution element).

Data reduction was performed in the standard way within the IRAF\footnote{IRAF
is distributed by the National Optical Astronomy Observatories, which are
operated by the Association of Universities for Research in Astronomy, Inc.,
under cooperative agreement with the National Science Foundation} environment.
Equivalent widths were measured by fitting Gaussian profiles to the absorption
lines, and will be reported in \citet{aoki00}.  The error for lines weaker than
60 m\AA, determined from the comparison of two measurements of lines which
appear on adjacent \'echelle orders, was about 4 m\AA \ and 6 m\AA \ in the
blue and red spectra, respectively. There is no systematic difference between
the equivalent widths of Fe {\small I} lines measured in this work and those by
\citet{norris96}, even though our $S/N$ is substantially higher.  

Additional spectra were obtained on 1998 August 11 and 2000 January 26, the
former using UCLES, and the latter with the Utrecht \'echelle spectrograph
(UES) on the William Herschel Telescope (WHT).  Both have lower S/N than is
necessary for an abundance analysis, the sole aim being to measure radial
velocities.  In each case, HD~140283 was also observed to provide a template
for cross-correlation.  That star has a similar metallicity but is free of the
CH blends that affect many lines in LP~625-44.  Radial velocities for LP~625-44
were obtained relative to HD~140283 by cross-correlation, and by measuring the
radial velocity of HD~140283 from the central wavelengths of 175 (1998) and
122 (2000) unblended lines. Error estimates were based on the variation in
velocity from different \'echelle orders and from the standard error in the
measurement of HD~140283. The heliocentric values, which extend those presented
by \citet{norris97a}, are: HJD 2451037.00: $v_{\rm rad}$~=~33.5$\;\pm\;$0.2
(1$\sigma$) km s$^{-1}$; and HJD 2451569.80: $v_{\rm rad}$~=~30.0$\;\pm\;$0.3
(1$\sigma$) km s$^{-1}$.  \citet{ryan99} estimated external errors of
0.3~km~s$^{-1}$ for a similar procedure; this has been added to the internal
errors for Fig.~\ref{fig:rv}.  The data confirm that LP~625-44 is a binary with
a period of at least 12 years.

\section{Abundance Analysis and Results}\label{sec:ana}

In the region near the Pb {\small I} line at 4058\AA\ , line-blending is so
severe that the Pb abundance was derived by spectrum synthesis.  This method
was also applied to lines which are affected by blending and/or hyperfine
splitting. The standard analysis, based on the equivalent widths, was applied
to single (unblended) lines.

The abundance analysis used model atmospheres in the ATLAS9 grid of
\citet{kurucz93a}. We adopted an effective temperature $T_{\rm eff}=5500$K,
determined by \citet{norris97a} from the $R-I$ color.  This color is not
severely affected by strong carbon and nitrogen features in stars of this
temperature and abundance \citep{aoki00}. Surface gravity ($\log g$),
metallicity, and microturbulent velocity ($\xi$), were re-determined in the
present work.  The surface gravity was obtained from the ionization balance
between Fe {\small I} and Fe {\small II}, the metallicity was estimated from
the abundance analysis of those lines, and  $\xi$ was determined from the Fe
{\small I} lines by demanding no dependence of the derived abundance on
equivalent widths.  The results are: $\log g=2.8$, $\xi=1.6$ km s$^{-1}$, and
[Fe/H] $=-2.72$. The agreement with the results of \citet{norris97a} is good,
with the exception of the microturbulent velocity, for which they derived 1.0
km s$^{-1}$.

Pb lines are difficult to measure in optical stellar spectra. Even for the sun,
only four Pb {\small I} lines (3639 \AA\ , 3683 \AA\ , 3739 \AA\ , and 4057
\AA) have been studied in the visual region. From these, \citet{youssef89}
determined the abundance $\log\epsilon_{\rm Pb}=2.0$, which agrees fairly well
with meteoritic measurements \citep[$\log\epsilon_{\rm
Pb}=2.06;$][]{grevesse96}. 

Our spectra covered the lines at 3739.9\AA \ and 4057.8\AA \ , listed in Table
\ref{tab:line}. They are expected to be weak, and no clear absorption
appears in the solar spectrum.  However, Pb {\small I} 4057.8\AA \ was
clearly detected in LP~625-44, as shown in the upper panel of
Fig.~\ref{fig:pb}, where the synthetic spectra fitted to the observed
data are also shown.  In spite of the presence of other lines, the
contribution of Pb {\small I} is remarkable.
To check possible contamination of the Pb region by other elements, we examined
the spectrum of HD~140283, a very metal-poor subgiant ($T_{\rm eff}=5750$~K,
$\log g=3.4$ and [Fe/H] $=-$2.54, Ryan et al. 1996).  We found no distinct
absorption feature at 4057.8\AA \ \citep[see Fig.~1b in ][]{norris96}. As a
further check for contamination due to CH and CN, the observed and synthetic
spectra of CS~22957-027 are shown in the lower panel of Fig.~\ref{fig:pb}.
\citet{norris97b} showed that this very metal-poor giant ($T_{\rm eff}=4850$K,
$\log g=1.9$ and [Fe/H] $=-3.38$) has very large excesses of $^{12}$C, $^{13}$C
and N but no excess of heavy elements. This spectrum indicates that the
absorption feature at 4057.8\AA \ in LP~625-44 is {\it not} due to CH and CN
lines.

To check our procedure for the determination of the Pb abundance, we also
analyzed the solar spectrum \citep{kurucz93b} using a solar photospheric model
\citep{holweger74}. Our result agrees very well with that of
\citet{youssef89} for Pb {\small I} 3683\AA,    which is the clearest
Pb {\small I} line in the optical range. This demonstrates
the reliability of the basic data (e.g., partition functions) and the software
used in our analysis. (Line contamination is so severe at 4057.8\AA \ in the
solar spectrum that the exact abundance cannot be derived from this line.)

An abundance ratio [Pb/Fe] $= 2.65$ was derived for LP~625-44
from a comparison between the synthetic spectra and the observed one.  The
other Pb {\small I} line covered by our spectrum is at 3739.9\AA, but there is
no distinct feature at this wavelength. Hence, we derive an upper limit on the
abundance ratio [Pb/Fe] $<+3.2$ from this non-detection, which supports
the Pb {\small I} 4057.8\AA \ result ([Pb/Fe]=2.65).  This upper limit is
important, as in the next section we show that this Pb abundance is {\it
lower} than predicted by some models of $s$-process nucleosynthesis in very
metal-poor AGB stars.

Our derived abundances for the heavy elements are similar to the results
presented by \citet{norris97a}.  Abundances of Er, Tm and Hf, not previously
known in this star, could also be determined due to the better quality of the
new spectra.  All new results are given in Table \ref{tab:res}. The line data
used in the analysis will be compiled in \citet{aoki00}.

Errors in our estimated abundances were obtained as follows.  Errors arising
from uncertainties in the atmospheric parameters were evaluated by adding in
quadrature the individual errors on the parameters -- $\Delta T_{\rm
eff}=100$K, $\Delta \log g=0.3$, and $\Delta \xi=0.5$km s$^{-1}$. The internal
errors were estimated by assuming the random error in the equivalent width
measurements to be 4 m\AA \ (and 6 m\AA \ for Ba {\small II} in the red region;
see \S \ref{sec:obs}), and taking the random error in less-certain $gf$ values
to be 0.1 dex.

\section{Discussion and Concluding Remarks}\label{sec:disc}

Fig.~\ref{fig:abundance} presents derived abundances as a function of
atomic species for LP~625-44. The thick solid line indicates the
abundance pattern of the main {\it s}-process component determined by
\citet{arlandini99}, while the thin line indicates the {\it r}-process
component.  The dotted line is the total solar abundance adopted by
\citet{arlandini99}.  We see good agreement between the observed
abundances of elements heavier than Ba with the scaled {\it s}-process
component. This fact, found by
\citet{norris97a} for Ba to Dy, is now extended to heavier elements and made
even more compelling.  The excesses of these elements, and their {\it
s}-process nature, are interpreted as a result of the transfer of material rich
in {\it s}-process elements across a binary system including an AGB star.  Our
new radial velocity measurements confirm the binarity and strengthen this
interpretation.  Since the excess of heavy elements is very large (e.g.,
[Ba/Fe]=2.7), the material from the AGB star should dominate the surface
abundances of LP~625-44.  Thus, the relative abundances of the heavy elements
in this star should provide an almost pure representation of the
nucleosynthesis products of the previously existing very metal-poor ([Fe/H]
$=-2.7$) AGB companion.

With the adoption of this interpretation the abundances in LP~625-44 can be
compared with theoretical predictions of nucleosynthesis in AGB stars.
\citet{gallino98} and \citet{busso99} showed that, at low abundance,
the metallicity effect on {\it s}-process yields favors the production of
heavier elements. As found in Fig.~\ref{fig:abundance}, the Sr-Zr enhancement
relative to the solar {\it s}-component is much smaller than that of heavy
elements (Ba-Hf) in LP~625-44, a result in qualitative agreement with the
expected metallicity dependence.

The metallicity effect is essentially due to the level of neutron exposure,
which is expected to be higher at lower metallicity (see \S
\ref{sec:intro}). Higher neutron exposure necessarily requires larger
production of the heaviest s-process element, Pb, in very metal-poor stars. 
\citet{busso99} explicitly showed the metallicity effect on the enhancement
factors of {\it s}-process elements relative to solar abundances for $-3.2 < $
[Fe/H] $ < 0.4$ in their Figure 12, where the enhancement factor of Pb is
larger by about two orders of magnitude than that of Ba at [Fe/H] $= -2.7$.
However, the enhancement of Pb in LP~625-44 ([Pb/Fe] = 2.65) is {\it nearly the
same} as that of Ba ([Ba/Fe] = 2.74).  If the observed Pb abundance of
LP~625-44 generally represents the yields from very metal-poor AGB stars, their
models of nucleosynthesis in AGB stars may overestimate its production by
about two orders of magnitude at very low metallicity.

This conflict might be resolved by tuning the models of \citet{gallino98} or
\citet{busso99}.  For instance, the neutron flux can be changed by modifying
the extension or chemical profile of the $^{13}$C-pocket, which is a free
parameter in their models. Another parameter is the mass of the AGB star, upon
which the number of thermal pulses (and hence episodes of neutron exposure) is
strongly dependent.

Another possibility is that the {\it s}-process production mechanism in very
metal-poor (e.g., [Fe/H]$<-2.5$) AGB stars is quite different from that in more
metal-rich stars. The calculation of low-mass stellar evolution in
metal-deficient stars by \citet{fujimoto00} showed that hydrogen mixing occurs
during the helium shell flash for 1$-$3.5 M$_{\odot}$ stars with [Fe/H]$<-2.5$
(their case II'), contrary to the situation for stars with higher metallicity
(their case IV). Their result suggests that the production of $^{13}$C, and
subsequent {\it s}-process nucleosynthesis, is possible in the helium
convective region during thermal pulses in these very metal-poor stars.

Our detection of Pb in the very metal-poor, carbon- and {\it s}-process-rich
star, LP~625-44 provides a strong constraint on models of nucleosynthesis in
AGB stars.  The observed abundance patterns for heavier elements (Ba-Pb) agree
well with the solar main {\it s}-process component, rather than with
nucleosynthesis models for very metal-poor AGB stars. Further observation of
{\it s}-process elements, including Pb, for objects similar to LP~625-44, and
revisions of the theoretical approach to the nucleosynthesis in very metal-poor
environments, will impact on our understanding of the evolution of low- and
intermediate-mass stars, as well as of the enrichment of heavy elements in the
early Galaxy.  In this context, we note that we have also measured Pb in the
star LP~706-7, which is similar to LP~625-44 in many respects
\citep{norris97a}. That object, whose {\it s}-process abundances nevertheless
differ from those of LP~625-44, will be discussed separately in a future paper.

\acknowledgments

We are grateful to the Director and staff of the Anglo-Australian Observatory,
and the Australian Time Allocation Committee for providing the observational
facilities used in this study. W.A. would like to acknowledge fruitful
discussions with T. Kajino.  T.C.B.  acknowledges partial support of this work
from grant AST 95-29454 awarded by the (US) National Science Foundation.

\newpage
\begin{table}
\
\caption[]{Pb {\small I} lines and Results for LP~625-44}
\label{tab:line}
\begin{tabular}{cccc}
\noalign{\smallskip}
\hline
\noalign{\smallskip}
Wavelength (\AA) & $\chi$ (eV) & $\log gf$ & [Pb/Fe] \\
\noalign{\smallskip}
\hline
\noalign{\smallskip}
3739.940 & 2.66 & $-0.12$ & $<+3.2$ \\
4057.815 & 1.32 & $-0.20$ & +2.7 \\
\noalign{\smallskip}
\hline
\end{tabular}
\end{table}
\begin{table}
\caption[]{Heavy Element Abundances for LP~625-44}
\label{tab:res}
\begin{tabular}{ccccc}
\noalign{\smallskip}
\hline
\noalign{\smallskip}
Element \hspace{2cm} & [X/Fe] & $\log\epsilon_{\rm el}$ & n & $\sigma$ \\
\noalign{\smallskip}
\hline
Fe I ([Fe/H]) \dotfill  & $-$2.71 & 4.78 & 34 & 0.13 \\
Fe II ([Fe/H]) \dotfill & $-$2.70 & 4.79 & 3  & 0.18 \\
Sr II \dotfill          & +1.15   & 1.37 & 3  & 0.16 \\
Y II  \dotfill          & +0.92   & 0.45 & 2  & 0.12 \\
Zr II \dotfill          & +1.31   & 1.22 & 4  & 0.12 \\
Ba II \dotfill          & +2.74   & 2.26 & 3  & 0.20 \\
La II \dotfill          & +2.50   & 1.02 & 5  & 0.13 \\
Ce II \dotfill          & +2.27   & 1.20 & 26 & 0.12 \\
Pr II \dotfill          & +2.45   & 0.55 & 5  & 0.12 \\
Nd II \dotfill          & +2.22   & 1.00 & 16 & 0.12 \\
Sm II \dotfill          & +2.20   & 0.48 & 16 & 0.12 \\
Eu II \dotfill          & +1.97   & $-$0.2 & 2  & 0.20 \\
Gd II \dotfill          & +2.31   & 0.70 & 6  & 0.13 \\
Dy II \dotfill          & +1.98   & 0.1  & 4  & 0.2 \\
Er II \dotfill          & +2.04   & 0.3  & 2  & 0.2 \\
Tm II \dotfill          & +1.96   & $-0.6$ & 1  & 0.2 \\
Hf II \dotfill          & +2.76   & 0.8 & 2  & 0.2 \\
Pb I  \dotfill          & +2.65   & 2.0  & 1  & 0.2 \\
\noalign{\smallskip}
\hline
\end{tabular}
\end{table}
\begin{figure}
\caption[]{
\label{fig:rv}
Radial velocity as a function of Julian Day for LP~625-44. 
}
\end{figure}

\begin{figure}
\caption[]{  Comparison of the observed (dots) and synthetic 
(lines) spectra near Pb I 4057.8\AA. In the upper panel,
LP~625-44 is shown along with the four synthetic spectra for
[Pb/Fe] = 0.0, 2.35, 2.65 and 2.95. The atomic and molecular species that
strongly contribute to the absorption are also labelled.  For
comparison, and for a check on possible contamination from
molecular lines, the spectrum of CS~22957-027, and the synthetic spectrum for
[Pb/Fe] $= 0.0$ are shown in the lower panel. }
\label{fig:pb}
\end{figure}

\begin{figure}
\caption[]{
The abundances of heavy elements as a function of atomic species for
LP~625-44. The thick solid line indicates
the main {\it s}-process component determined
by \citet{arlandini99} using models of 1.5 M$_{\odot}$ and 3 M$_{\odot}$
AGB stars at $Z=\frac{1}{2}Z_{\odot}$, while the thin line indicates the {\it
r}-process component derived by subtraction of the {\it s}-process component
from the solar abundances. The exception is the {\it r}-process component of
Pb, for which the value determined by \citet{kappeler89} is adopted. The dotted
line represents the total solar abundance adopted in \citet{arlandini99}. All
abundance patterns are normalized to the observed Ba abundance of LP~625-44.  }
\label{fig:abundance}
\end{figure}

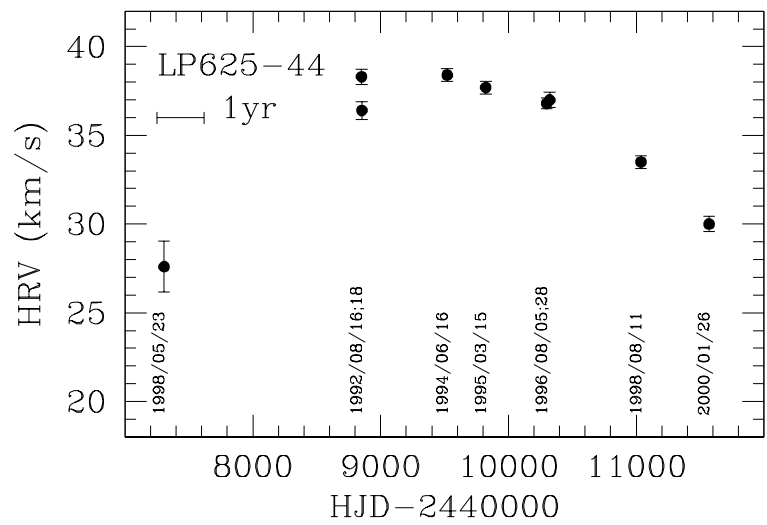
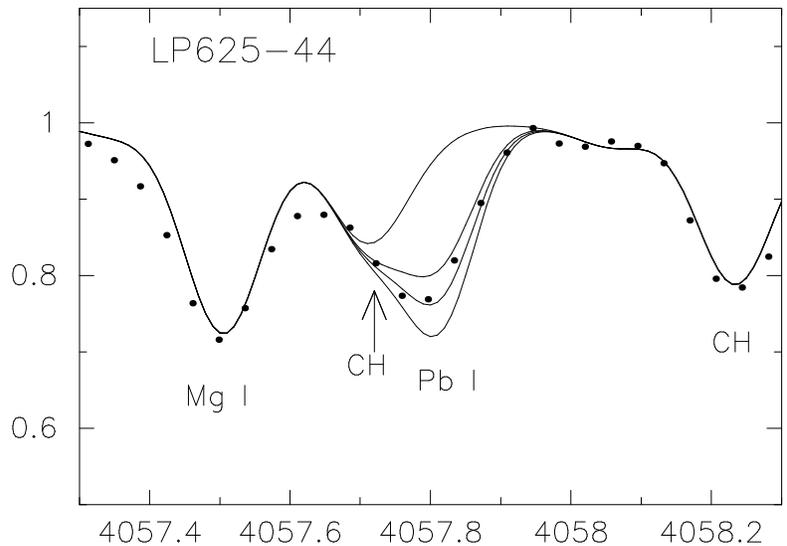
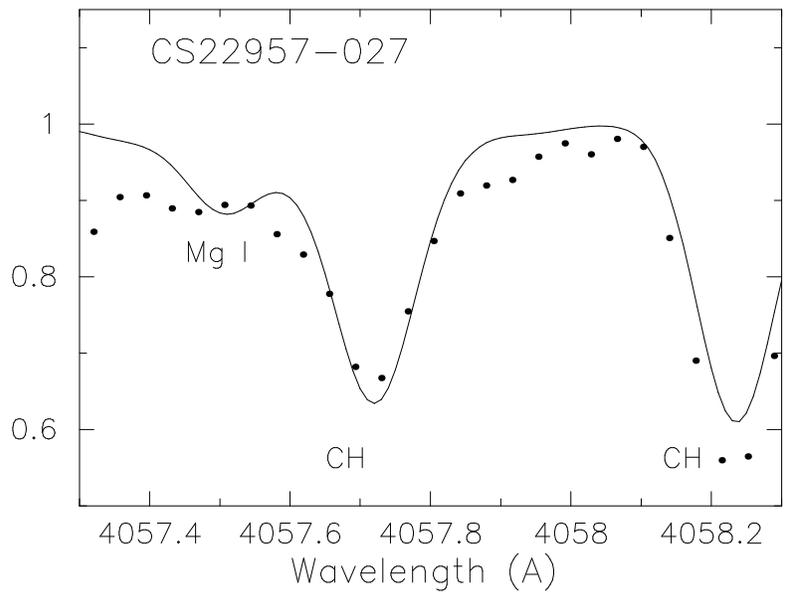
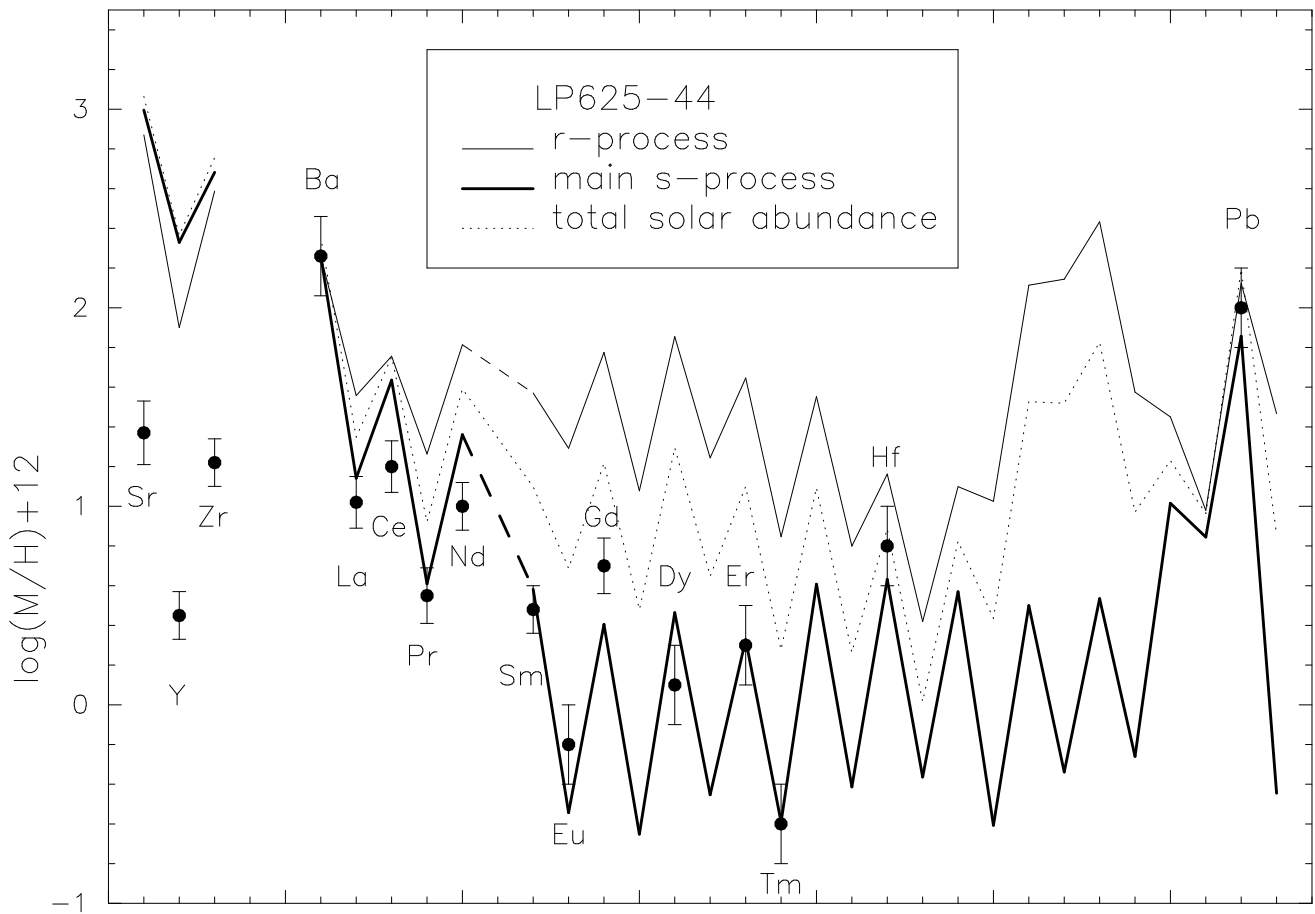
\end{document}